\documentclass[aps,twocolumn,showpacs,groupedaddress,nofootinbib]{revtex4}  % for review and submission
\usepackage{graphics,graphicx}  % needed for figures
\usepackage{dcolumn}   % needed for some tables
\usepackage{bm}        % for math
\usepackage{amssymb}   % for math
\usepackage{bigints}
\usepackage{xcolor}
\def\be{\begin{equation}}
\def\ee{\end{equation}}
\def\bea{\begin{eqnarray}}
\def\eea{\end{eqnarray}}

\begin{document}

% The following information is for internal review, please remove them for submission
\widetext

\title{The Hydrogen Bond of QCD
%Tetraquarks with Hidden and Open Double Heavy Flavour in the Born-Oppenheimer Approximation
}
\author{Luciano Maiani}
\author{Antonio D. Polosa}
\author{Veronica Riquer}
\affiliation{Dipartimento di Fisica and INFN,  Sapienza  Universit\`a di Roma, Piazzale Aldo Moro 2, I-00185 Roma, Italy.}
\email{luciano.maiani@roma1.infn.it}
%\author{Antonio D. Polosa}
%\affiliation{Dipartimento di Fisica and INFN,  Sapienza  Universit\`a di Roma, Piazzale Aldo Moro 2, I-00185 Roma, Italy}
\email{antonio.polosa@roma1.infn.it}
\email{veronica.riquer@cern.ch}
%\affiliation{Dipartimento di Fisica and INFN,  Sapienza  Universit\`a di Roma, Piazzale Aldo Moro 2, I-00185 Roma, Italy.} 

\date{\today}

\begin{abstract}
Using the Born-Oppenheimer approximation, we show that exotic resonances, $X$ and $Z$, may emerge as QCD molecular objects made of colored two-quark lumps, states with heavy-light diquarks spatially separated from antidiquarks. 
With the same method we confirm that doubly heavy tetraquarks  are stable against strong decays. 
Tetraquarks described here provide a new picture of exotic hadrons, as formed by the QCD analog of the hydrogen bond of molecular physics. 
%Color forces in multiquark hadrons allow, under specific conditions, states with heavy-light diquarks spatially separated from antidiquarks, with strong phenomenological consequences.  The same method used to reach this conclusion shows that doubly heavy tetraquarks are indeed stable against strong decays. Exotic resonances like $X$ and $Z$ emerge as molecular objects made of colored two-quark lumps which can be found in symmetric and anti-symmetric color configurations, held together in color neutral hadrons by the QCD analog of the hydrogen bond. 
\end{abstract}

\pacs{12.40.Yx, 12.39.-x, 14.40.Lb}
\maketitle
% sections are not used for PRL papers
{\bf \emph{Introduction.}}\label{intro}
In this letter we present a description of tetraquarks%particles
~\cite{book,Esposito:2016noz,Maiani:2004vq} in terms of {\it color molecules}: two lumps of two-quark (colored atoms) held together  by color forces. The variety of tetraquarks decribed here identifies a new way of looking at multiquark hadrons, as formed by the QCD analog of the hydrogen bond of molecular physics. 

We restrict to heavy-light systems, $Q\bar Q q \bar q$ or $QQ\bar q\bar q$, and apply  the Born-Oppenheimer (BO) approximation,  see {\it e.g.}~\cite{weinbergQM}, the method used for the hydrogen molecule, see~\cite{pauling}. 
The method consists in solving the eigenvalue problem for the light particles with fixed coordinates of the heavy ones, ${\bm x}_A, {\bm x}_B$, and then solve the Schr\"odinger equation of the heavy particles in the BO potential
\be
V_{\rm BO}({\bm x}_A, {\bm x}_B)= V({\bm x}_A, {\bm x}_B) + {\cal E}({\bm x}_A, {\bm x}_B)
\ee
$V({\bm x}_A, {\bm x}_B) $ is the interaction between the heavy particles, {\it e.g.} the electrostatic repulsion, and ${\cal E}({\bm x}_A, {\bm x}_B)$ is the lowest energy eigenvalue of the light particles at fixed heavy particles coordinates. The approximation improves with $m_q/M_Q\to 0$.

The application of the Born-Oppenheimer method to doubly heavy tetraquarks in lattice QCD has been suggested recently in~\cite{bicudo,lebed}, both for hidden flavor tetraquarks, $[cq][\bar c\bar q]$, {\it i.e.} the exotic resonances $X, Z$~\cite{Maiani:2004vq,Maiani:2014aja,Maiani:2017kyi,Ali:2017wsf}, and for double beauty open flavor tetraquarks, $bb\bar q\bar q$, introduced in~\cite{Esposito:2013fma,Guerrieri:2014nxa} and, more recently, studied in~\cite{Karliner:2017qjm,Eichten:2017ffp,Eichten:2017ual,Luo:2017eub}.  

%Besides coordinates, we have to fix the color of the heavy quarks. 
We fix the $Q\bar Q$ pair to be in color ${\bm 8}$
%~\footnote{With $Q\bar Q$ in color singlet, the interaction with the light quark pair would be mediated by color singlet exchanges, $\pi,~\rho,$~etc. and we would fall, eventually, in the hadron molecule case~\cite{braaten}, to be treated with the methods of nuclear physics.} 
 and we consider both possibilites, ${\bar{\bm 3}}$~{and ${\bm 6}$, for $QQ$. 
 Had we taken $Q\bar Q$ in color singlet, the interaction with the light quark pair would be mediated by color singlet exchanges, as in the hadroquarkonium model proposed in~\cite{Dubynskiy:2008mq}.

For hidden flavour tetraquarks, we obtain color repulsion within the heavy $Q\bar Q$ and the light $q\bar q$ quark pairs, and mutual attraction between heavy and light quarks or antiquarks. Thus, in the $[Qq]-[\bar Q\bar q]$ color singlet molecule, repulsions and attractions among constituents are distributed in the same way as for protons and electrons in the hydrogen molecule. 
Assuming one-gluon exchange forces, Fig~\ref{uno}(a)  describes a configuration of a tight $Q\bar Q$ similar to the  ``quarkonium adjoint meson" discussed  in~\cite{braatenBO}, see also~\cite{Brambilla:2017uyf}.
Increasing the repulsion between light quarks beyond the naive one-gluon exchange force, we obtain a configuration of the potential which  separates the diquarks  from each other, Fig.~1(b), as envisaged in~\cite{Selem:2006nd},  with   the phenomenological implications discussed in~\cite{Maiani:2017kyi}  and~\cite{Esposito:2018cwh}. The most compelling one  is that decays of $X,Z$ particles into quarkonia+mesons are suppressed with respect to decays into open charm mesons: the tunneling of heavy quark pairs through the barrier gets a larger suppression factor.   At difference from what was done originally in~\cite{Maiani:2004vq, Maiani:2014aja, Maiani:2017kyi}, %assuming a  mutual repulsion between the light and heavy quarks, 
the two lumps of two-quark states $Qq+\bar Q \bar q$ are found  in a superposition of  diquark-antidiquark in the $\bar{\bm 3}\otimes \bm 3$   and  $\bm 6 \otimes\bar{\bm 6}$ color configurations.

The two light particles are not equal and there are two different heavy-light orbitals: %with one light quark coupled to one heavy particle. 
%Therefore 
in addition to  $Qq+\bar Q \bar q$, we  examine the  $Q\bar q+ \bar Q q$ case. 
In  the latter,  $Q\bar q$ and $\bar Q q$ orbitals have a %large color singlet component
color octet component. As we dhall see, however, at large separations between heavy quarks  the lowest state will correspond to a pair of color singlet charmed mesons. 
A minimum of the BO potential is not guaranteed. If there is such a minimum, as in Fig.~\ref{due}(a), it would correspond to a configuration similar to the quarkonium adjoint meson of the previous case. 
If repulsion in the $q\bar q$ pair prevails, there is no minimum at all, Fig.~\ref{due}(b). 

The BO potential for $(QQ)_{\bar{\bm 3}}$ is presented in Fig.~\ref{qq3}. The unperturbed orbitals correspond to $Q\bar q$ and $\bar Q q$. Forces among constituents are all attractive and the potential vanishes at large $QQ$ separation.   This allows a new, independent estimate of the extra binding of $QQ$. %with respect to the naive constituent quark model. 
We confirm the result obtained in~\cite{Karliner:2017qjm,Eichten:2017ffp,Luo:2017eub}  with different variants of the naive constituent quark model, that the lowest $bb$ tetraquark and possibly $bc$ are stable under strong decays, while $cc$ is borderline, see Tab.~\ref{tab}.

$(QQ)_{{\bm 6}}$ repel each other. However, with the constraint of an overall color singlet, we find both attractive and repulsive forces and the BO potential may admit a second $QQ$ tetraquark. With the perturbative one-gluon-exchange couplings, a shallow bound state is indeed found.

In conclusion, the BO approximation, even with the limitations of our perturbative treatment, gives a new insight on the tetraquark structure and provides new opportunities in the intricate field of exotic resonances properties. We hope that our approach may be the basis of further investigations on the internal structure of multiquark hadrons and the phenomenology of their decays.  Non-perturbative investigations along these lines should be provided by lattice QCD (see for example~\cite{bicudo}), following the growing interest shown for doubly heavy tetraquarks~\cite{latts}.

 The picture of diquark-antidiquark states segregated in space by a potential barrier is compatible with the existence of charged partners of the $X^0(3872)$ to be found in $X^\pm\to \rho^\pm\,J/\psi$ final states, with  branching fractions considerably smaller than in the neutral channel.  This requires to push way further on the available experimental bounds.   It also gives an independent thrust to the idea of stable $bb\bar q\bar q$ tetraquarks, still awaiting an experimental confirmation.

%@@@@@@@@@@@@@@@@@@@@@@@@@@@@@@@@@@@@@@@
{\bf \emph{Hidden Charm.}}\label{intro}
%We consider here the hidden charm tetraquark, $c q\bar c\bar q$. %According to the BO approximation, we need to fix the coordinates and the other quantum numbers of the heavy pair. We neglect spin altogether, we fix the flavour of $q$ and $\bar q$ ({\it e.g.} $q=u,~\bar q=\bar u$). 
We indicate with ${\bm x}_A$ and  ${\bm x}_B$ the coordinates of $c$ and $\bar c$, and  ${\bm x}_{1,2}$ the coordinates of $q$ and $\bar q$. Both $c\bar c$ and $q\bar q$ are taken in the $\bm 8$ color representation. 

Suppressing coordinates
$
T=(\bar c \lambda^a c)(\bar q \lambda^a q) \label{tetra}
$
with the sum over $a=1,\dots,8$ understood. 

If we restrict to one-gluon exchange we find the interactions between the different pairs
in terms of the quadratic Casimir operators 
 \be 
 \lambda_{q_1 q_2}({\bm R})=\alpha_S\frac{1}{2}\Big(C_2({\bm R})-C_2(\bm q_1)-C_2(\bm q_2)\Big)
 \label{casimir}
 \ee
$\bm q_{1,2}$ are the $\bm 3$ or $\bar{\bm 3}$ irreducible representations of the color group depending on wether $q_{1,2}$ are quarks or antiquarks, and ${\bm  R}$ is the color representation of the $q_1q_2$ pair~\footnote{We recall the results: $C_2({\bm 1})=0$, $C_2({\bm R})=C_2({\bar{\bm R}})$, $C_2 ({\bm 3})=4/3$, $C_2({\bm 6})=10/3$, $C_2({\bm 8})=3$.}. 

If we find the pair $q_1 q_2$ in the tetraquark $T(q_i q_jq_kq_l)$  in a superposition of two SU(3)$_c$ representations with amplitudes $a$ and $b$
\be
T= a |(q_1 q_2)_{{\bm  R}_1}\cdots\rangle_{{\bm 1}}+b |(q_1 q_2)_{{\bm  R}_2}\cdots\rangle_{{\bm 1}}
\ee
%\be
%|q_1 q_2\rangle= a |(q_1 q_2)_{{\bm  R}_1}\rangle+b |(q_1 q_2)_{{\bm  R}_2}\rangle
%\ee
then we use 
\be
\lambda_{q_1q_2}=a^2\lambda_{q_1 q_2}({\bm R_1})+b^2\lambda_{q_1 q_2}({\bm R_2})
\ee

Since both $c\bar c$ and $q\bar q$ are in color octet we have
$\lambda_{c\bar c}=\lambda_{q\bar q}= +1/6\,  \alpha_S$.  The couplings of the other pairs are found using the Fierz rearrangement formulae for SU(3)$_c$ to bring the desired pair in the same quark bilinear. 
We get %(see Appendix~\ref{fierz}):
 \be
 \lambda_{cq}=\lambda_{\bar c\bar q}= -\frac{1}{3}\alpha_S\quad %\label{attrcq}\\
\lambda_{c\bar q}=\lambda_{\bar c q}= -\frac{7}{6}\alpha_S\label{attrcqbar}
 \ee

The pattern of repulsions and attractions %mplied by previous formulae 
in (\ref{attrcqbar}) is the same as in the hydrogen molecule, substituting electrons with light and protons with heavy quarks. 
We take a perturbative approach similar to the one in the $H_2$ case~\cite{pauling}. 
For fixed coordinates of the heavy particles, ${\bm x}_A$ and ${\bm x}_B$, we describe the unperturbed state as the product of two {\it orbitals},  {\it i.e.} the wave functions of the bound states of one heavy and one light particle around ${\bm x}_A$ and ${\bm x}_B$, and treat the interactions not included in the orbitals as perturbations. 

Two subcases are allowed: $i)$  $cq$ (and $\bar c \bar q$) or $ii)$ $c\bar q$ (and $\bar c q$).  

{\bf \emph{The ${\bm c}{\bm q}$ orbital.}}
In the $H_2$ molecule, the orbital is just the hydrogen atom wave function in the ground state. In our case, we take the Coulombic interaction given by $\lambda_{cq}$ in (\ref{attrcqbar}) 
with the addition of a confining linear potential
\be
\label{orbitpot}
V_{cq}=-\frac{1}{3}\frac{ \alpha_S}{r}+kr + V_0
\ee

We assume a radial wave-function $R(r)$ of the form
\be
R(r) =\frac{A^{3/2}}{\sqrt{4\pi}}~e ^{-Ar}\label{radwf}
\ee
and determine $A$ by minimizing the Schroedinger  functional
\be
\langle H(A)\rangle=\frac{\Big(R(r),(-\frac{1}{2M_q}\Delta+V_{cq}-V_0)R(r)\Big)}{(R(r),R(r))}
\ee
We use a costituent light quark mass\footnote{For heavy quarks we take $M_c=1.67$~GeV, $M_b=5.0$~GeV~\cite{book,Maiani:2004vq}. } $M_q=0.31$~GeV estimated from the meson spectrum~\cite{book,Maiani:2004vq}, $\alpha_S=0.30$ at the charm mass scale and $k=0.15~{\rm GeV}^2$ from~\cite{lattice}. Another option is that $k$ follows the coefficient of the Coulombic force~\cite{Bali:2000gf}, which leads to $k=1/4\times 0.15$~GeV$^2$. We comment later on this alternative. 

We find
$
A=0.43~{\rm GeV}, \, \langle H\rangle_{\rm min} = 0.73 ~{\rm GeV} %+ V_0
%\label{parawf}
$.

We write the wave function of the $q\bar q$ state 
\be 
\Psi(1,2)=\psi(1)\phi(2)=R(|{\bm x}_1-{\bm x}_A|)R(|{\bm x}_2-{\bm x}_B|)
\ee
The unperturbed energy of $\Psi(1,2)$ is given by the quark constituent masses plus the energy of each orbital
$
E_0= 2(M_c+M_q + \langle H\rangle_{\rm min} +V_0)%\label{unperten}
$.

The perturbation Hamiltonian using the values for  $\lambda_{c\bar c}=\lambda_{q\bar q}$ and $\lambda_{c\bar q}=\lambda_{\bar c q}$ found above, is
\bea
H_{\rm pert}&=&-\frac{7}{6}\alpha_S \left(\frac{1}{|{\bm x}_1-{\bm x}_B|}+\frac{1}{|{\bm x}_2-{\bm x}_A|}\right) +\nonumber \\
&&+\frac{1}{6}\alpha_S\,\frac{1}{|{\bm x}_1-{\bm x}_2|}
\eea

 To first order in $H_{\rm pert}$ and with $r_{AB}=|{\bm x}_A-{\bm x}_B|$, the BO potential is 
\be
V_{\rm BO}(r_{AB})=+\frac{1}{6}\alpha_S\frac{1}{r_{AB}}+ \delta E %\langle \Psi(1,2)|H_{\rm pert}|\Psi(1,2)\rangle \nonumber \\
\ee
where %the first term in $V_{\rm BO}$ is the Coulomb interaction of $c\bar c$ and  
$\delta E= \big(\Psi(1,2),H_{\rm pert}\Psi(1,2)\big)$ evaluates to
\be
\delta E = -\frac{7}{6}\alpha_S \, 2 I_1(r_{AB})+
\frac{1}{6}\alpha_S\,  I_4(r_{AB})\label{vbopert}
\ee
%The first term in $V_{\rm BO}$ is the Coulomb interaction of $c\bar c$ and
The functions $I_{1,4}$ are given in~\cite{pauling} for hydrogen wave functions, and may be computed numerically for any  given  orbital~\eqref{radwf} 
\be
I_1(r_{AB})= \int d^3\xi\,|\psi(\xi)|^2 \, \frac{1}{|{\bm \xi}-{\bm x}_B|}\label{i1} 
\ee
where the vector $\bm \xi$ originates from $A$ and $|\bm x_B|=r_{AB}$. Similarly
\be
I_4(r_{AB})= \int d^3\xi d^3\eta\, |\psi(\xi)|^2\,|\phi(\eta)|^2 \frac{1}{|{\bm \xi}-{\bm \eta}|} \label{i4}
\ee

In addition, we take into account the confinement of the colored diquarks by adding a linearly rising potential % that starts at $r_{AB}\sim 10$~GeV$^{-1}\sim 2$~fm, where orbitals are separated:
determined by a string tension $k_T$ and the onset point, $R_0$
\bea
&& V_{\rm conf}(r)=k_T \times (r-R_0)\times \theta(r-R_0)\nonumber \\
&&V(r)=V_{\rm BO}(r) +V_{\rm conf}(r)\label{BOtot}
\eea
For orientation, we choose $R_0=10$~GeV$^{-1}$, greater than $2A^{-1}\sim 5$ GeV$^{-1}$, where the two orbitals start to separate~\footnote{$R_0$ should be considered a  free parameter, to be fixed on the phenomenology of the tetraquark,  as we discuss below.}. As for $k_T$, we note that
the tetraquark $T=|(\bar c c)_{\bm 8} (\bar q q)_{\bm 8}\rangle_{\bm 1}$ can be written as
\be
T=\sqrt{\frac{2}{3}} |(cq)_{\bar{\bm 3}} (\bar c \bar q)_{\bm 3}\rangle_{\bm 1}-\sqrt{\frac{1}{3}}|(cq)_{\bm 6} (\bar c \bar q)_{{\bar {\bm 6}}}\rangle_{\bm 1}\label{3&6}
\ee
At large distances the diquark-antidiquark system is a superposition of ${\bar {\bm 3}}\otimes {\bm 3}\to {\bm 1}$ and ${\bm 6}\otimes{\bar {\bm 6}} \to {\bm 1}$. 
The hypothesis of Casimir scaling of $k_T$~\cite{Bali:2000gf}  and (\ref{3&6}) would give
\be
k_T=\left(\frac{2}{3}+\frac{1}{3}~ \frac{C_2({\bm 6})}{C_2({\bm 3})}\right) k=1.5~k
\ee
However, as discussed in~\cite{Bali:2000gf}, gluon screening gives the ${\bm 6}$ diquark a component over the ${\bar{\bm 3}}$, which appears in the product ${\bm 6}\otimes {\bm 8}$, bringing $k_T$ closer to $k$. For simplicity, we adopt $k_T=k$.

The  potential $V(r)$ computed on the basis of Eqs.~(\ref{BOtot}) is given in Fig.~\ref{uno}(a). 
Also reported are the wave function and the eigenvalue obtained by solving numerically the radial Schr\"odinger 
equation~\cite{schroed}.

As it is customary for confined system like charmonia, we fix $V_0$ to reproduce the mass of the tetraquark, so the eigenvalue is not interesting. However, the eigenfunction gives us information on the internal configuration of the tetraquark. 
In Fig.~\ref{uno}(a), with one-gluon exchange couplings, a configuration with $c$ close to $\bar c$ and the light quarks around is obtained, much like the quarkonium adjoint meson described in~\cite{braatenBO}. 

Fig.~\ref{uno}(b) is obtained by increasing the repulsion in the $q\bar q$ interaction: $+1/6 \,\alpha_S\sim 0.11 \to 2.4$. 
The corresponding $c\bar c$ wave function clearly displays the separation of the diquark from the antidiquark. 
Had we used $k=1/4\times 0.15$~GeV$^2$ in Eq.~\eqref{orbitpot}, the required enhancement would be $+1/6 \,\alpha_S \to 3.3$.

The barrier that $c$ has to overcome to reach $\bar c$, apparent in Fig.~\ref{uno}(b), was suggested in~\cite{Maiani:2017kyi}, and further considered in~\cite{Esposito:2018cwh}, to explain the suppression of the $J/\psi+ \rho/\omega$ decay modes of $X(3872)$, otherwise favoured by phase space with respect to the $D D^*$ modes. 
Indeed, with the parameters in Fig.~\ref{uno}(b), we find $|R(0)|^2=10^{-3}$ 
~with respect to $|R(0)|^2=10^{-1}$ with the perturbative parameters of Fig.~\ref{uno}(a).

%@@@@@@@@@@@@@@@@@@@@@@@@@@
\begin{figure}[htb!]
\begin{minipage}[c]{4cm}
   \centering
 %\begin{center}
   \includegraphics[width=4truecm]{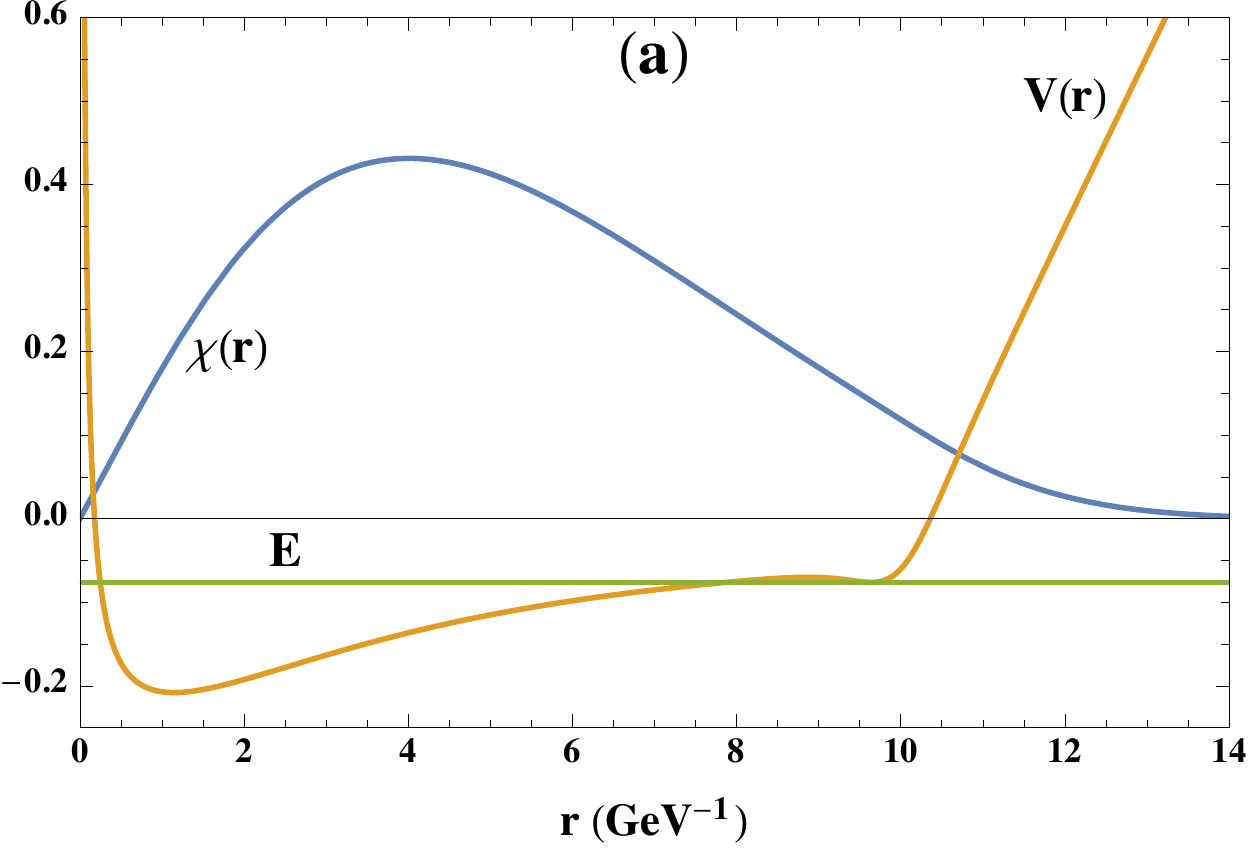}
    \end{minipage}%
% \end{center}
 \begin{minipage}[c]{5cm}
\centering
   \includegraphics[width=4truecm]{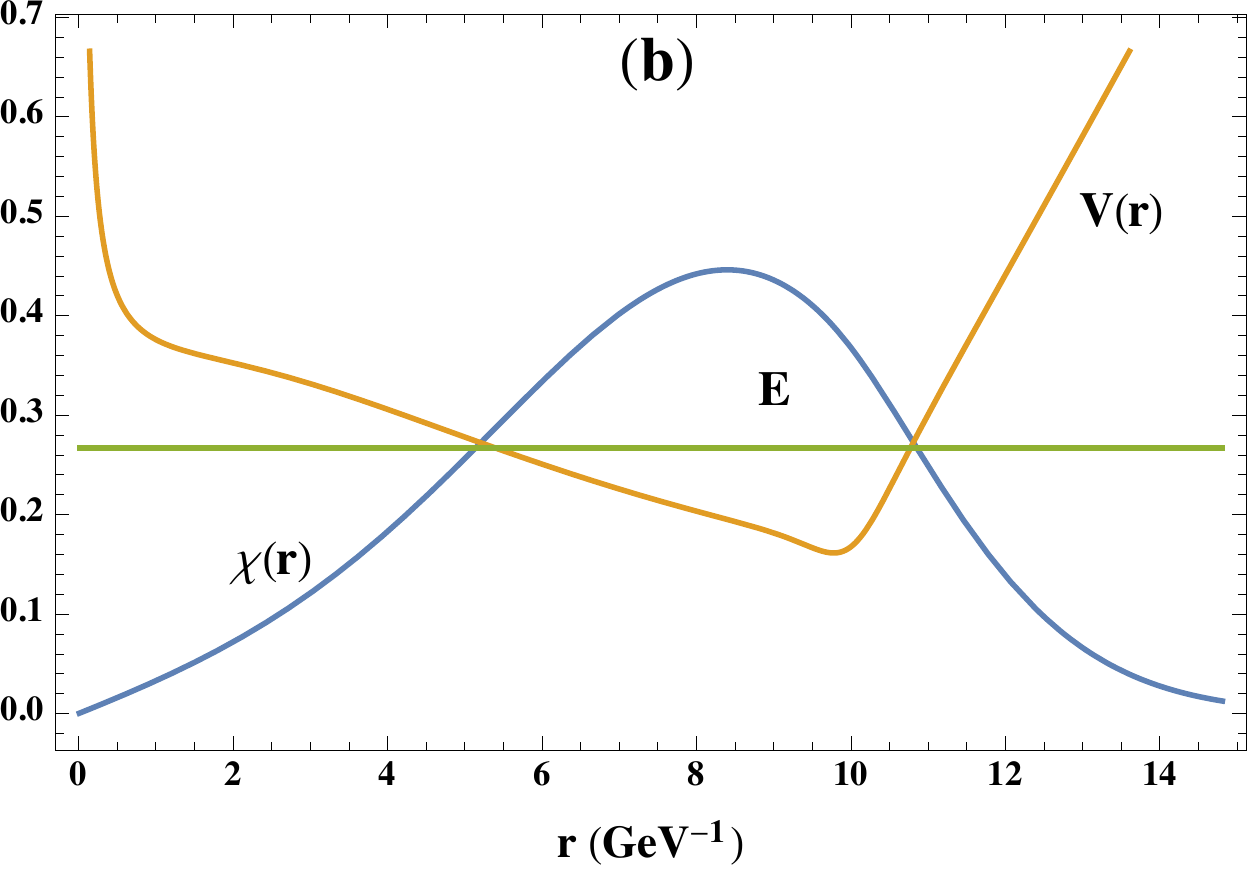}
 \end{minipage}%
\caption{\footnotesize (a) dominant $c\bar q$ and $\bar c q$ attraction + confinement; (b) dominant $q \bar q$ repulsion + confinement. Eigenfunction $\chi(r)=rR(r)$ and eigenvalue $E$ of the tetraquark in the fundamental state are shown. Diquarks are separated by a potential barrier and there are two different lenghts: $R_{qc}\sim 0.7-1$~fm and the total radius $R\sim 2.5$~fm~\cite{Maiani:2017kyi}. 
Here and in the following, on  the y-axes energies are in GeV and $\chi$ in arbitrary units.
\label{uno}}
\end{figure}

%@@@@@@@@@@@@@@@@@@@@@@@@@@@@@@

\begin{figure}[htb!]
\begin{minipage}[c]{4cm}
   \centering
 %\begin{center}
   \includegraphics[width=4.05truecm]{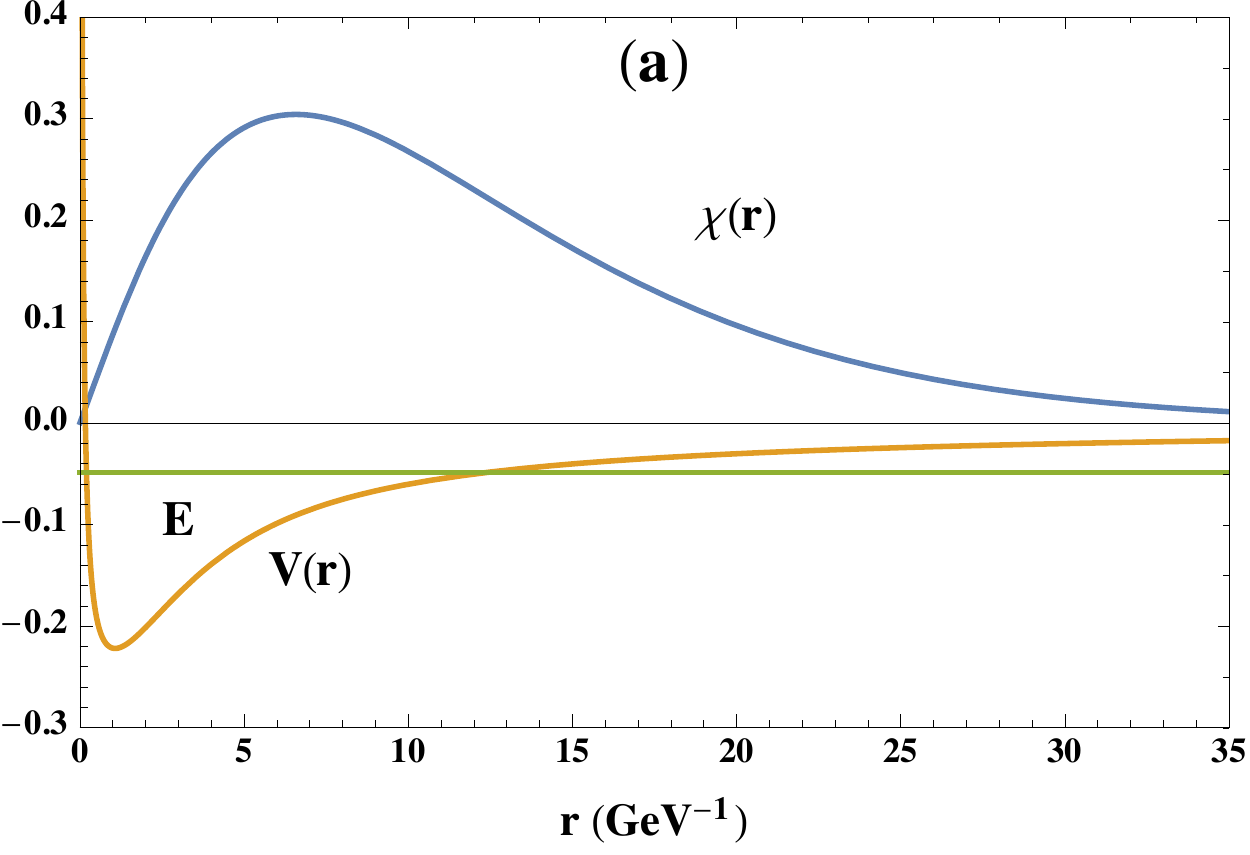}
    \end{minipage}%
% \end{center}
 \begin{minipage}[c]{5cm}
\centering
   \includegraphics[width=4truecm]{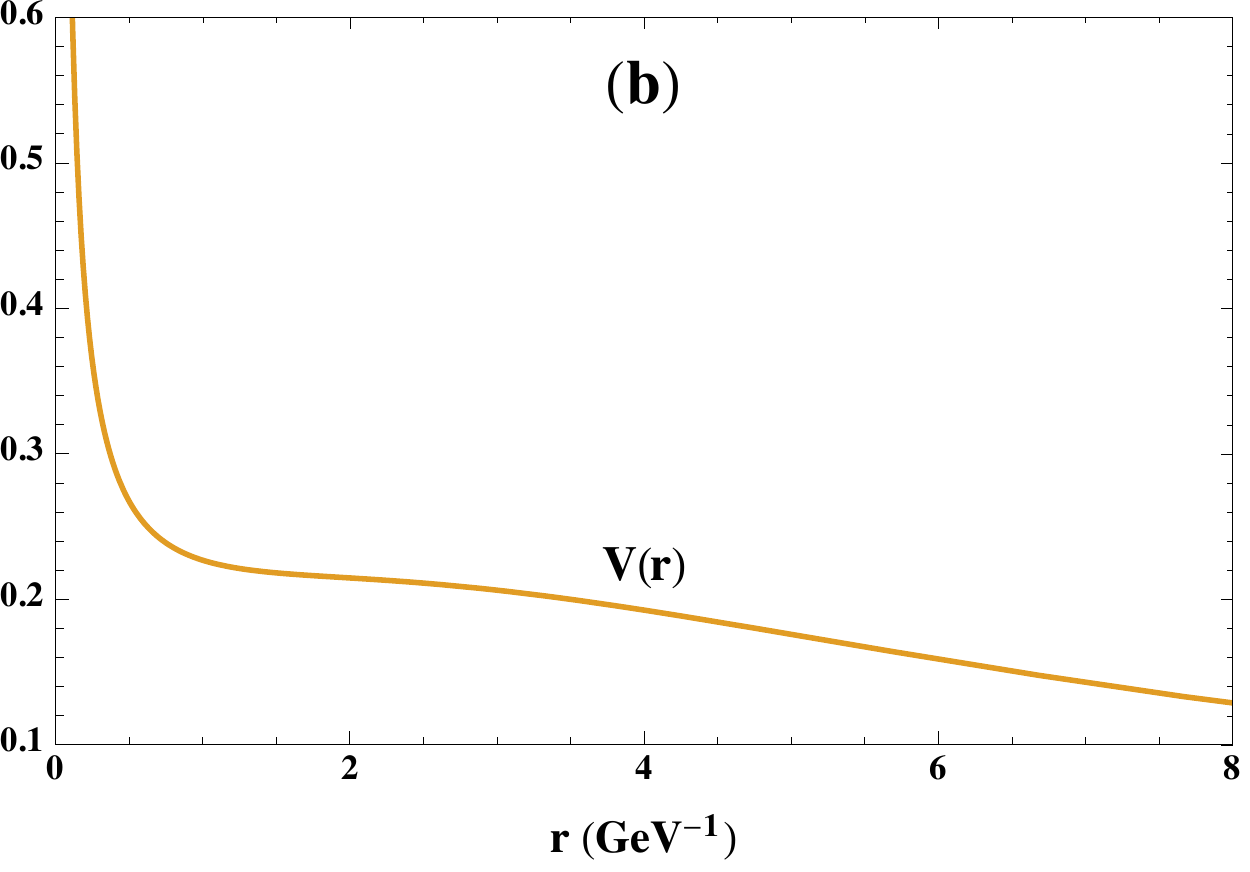}
 \end{minipage}%
\caption{\footnotesize {Born-Oppenheimer potential $V(r)$ vs. $R_{AB}$ for $c\bar q$ orbitals. Unit lenght: GeV$^{-1}\sim 0.2$~fm.   (a) using the perturbative parameters;%, $E=-17$~MeV;
~(b) with repulsion enhanced.}
}
\label{due}
\end{figure}

The tetraquark picture of $X(3872)$ and the related $Z(3900)$ and $Z(4020)$ have been originally formulated  in terms of pure ${\bar {\bm 3}}\otimes {\bm 3}$ diquark-antidiquark states~\cite{Maiani:2004vq, Maiani:2014aja, Maiani:2017kyi}. 
The ${\bm 6}\otimes{\bar {\bm 6}}$ component in (\ref{3&6}) results in the opposite sign of the $q\bar q$ hyperfine interactions vs the dominant $c q$ and $\bar c \bar q$ one, and it could be the reason why $X(3872)$ is lighter than $Z(3900)$.

%@@@@@@@@@@@@@@@@@@@@@@@@@@@@@@@@@@@@@@@@@
{\bf \emph{The ${\bm c}{\bar{\bm q}}$ orbital.}}
One obtains the new orbital by replacing  $-1/3~\alpha_S \to -7/6~\alpha_S$ in Eq.~(\ref{orbitpot}). Correspondingly
$
A=0.50~{\rm GeV},\, \langle H\rangle_{\rm min}  = 0.47 ~{\rm GeV}
$.
%The unperturbed  energy has the same form  as in  (\ref{unperten}), with the new value of $\langle H\rangle_{\rm min}+V_0 $. 
The perturbation Hamiltonian appropriate to this case is
\bea
H_{\rm pert}
&=&-\frac{1}{3}~\alpha_S \left(\frac{1}{|{\bm x}_1-{\bm x}_B|}+\frac{1}{|{\bm x}_2-{\bm x}_A|}\right) +\nonumber \\
&&+\frac{1}{6}~\alpha_S~\frac{1}{|{\bm x}_1-{\bm x}_2|} \label{pertcqbar}
\eea
and
\be
V_{\rm BO}=+\frac{1}{6}~\alpha_S ~\frac{1}{r_{AB}}+\delta E%\langle \Psi(1,2)|H_{\rm pert}|\Psi(1,2)\rangle\nonumber
\ee
The tetraquark state is 
\be
T=\sqrt{\frac{8}{9}}|(\bar c q)_{\bm 1}(\bar q c)_{\bm 1}\rangle_{\bm 1}-\frac{1}{\sqrt{9}}|(\bar c q)_{\bm 8}(\bar q c)_{\bm 8}\rangle_{\bm 1} \label{cqbar}
\ee
%and the lowest state will correspond, 
At large $|{\bm x}_A-{\bm x}_B|$ the lowest energy state
(two color singlet mesons) has to prevail, as concluded also in~\cite{Bali:2000gf} on the basis of the screening of octet charges due to gluons.

There is no confining potential and $V_{\rm BO}\to \langle H\rangle_{\rm min} +V_0$ for $r_{AB}\to \infty$. Including constituent quark masses, the energy of the state at $r_{AB}= \infty$ is  
$
E_\infty=2(M_c + M_q+ \langle H\rangle_{\rm min} +V_0)
$ 
and it must coincide with the mass of %the two orbitals corresponding to 
a pair of non-interacting charmed mesons, with spin-spin interaction subtracted. 
Therefore we impose
\be
\langle H\rangle_{\rm min} +V_0=0
\label{noconf}
\ee
A minimum of the BO potential is not guaranteed. If there is such a minimum, as in Fig.~\ref{due}(a), it would correspond to a configuration similar to the quarkonium adjoint meson in Fig.~\ref{uno}(a). %If repulsion prevails, there may be no minimum at all, Fig.~\ref{due}(b)
If repulsion is increased above the perturbative value,  {\it e.g.} changing $+1/6 ~\alpha_S\sim 0.11$ to a coupling $\geq 1$ in analogy with Fig.~\ref{uno}(b), the BO potential has no minimum at all, Fig.~\ref{due}(a).

 {\bf \emph{Double beauty tetraquarks: $\bm b\bm b$ in ${\bar{\bm 3}}$.}} 
%We start with the color antisymmetric state.
The lowest  energy state corresponds to $ bb$ in spin one and light antiquarks in spin and isospin zero. 
The tetraquark state $T=|(bb)_{\bar {\bm 3}}, (\bar q\bar q)_{ {\bm 3}} \rangle_{\bm 1}$ can be Fierz transformed into
\be
T=\sqrt{\frac{1}{3}}|(\bar q b)_{\bm 1},(\bar q b)_{\bm 1}\rangle _{\bm 1}-\sqrt{\frac{2}{3}}|(\bar q b)_{\bm 8},(\bar q b)_{\bm 8}\rangle _{\bm 1} \label{color3}
\ee
with all attractive couplings
\be
\lambda_{bb}=\lambda_{\bar q \bar q}=-\frac{2}{3}\alpha_S\quad %\label{bb3}\\
\lambda_{b\bar q}%=\lambda_{q \bar b}
=-\frac{1}{3}\alpha_S\label{bqbar3}
\ee

As in Eq.~\eqref{cqbar}, the ${\bm 8}$ charges are screend by gluons, so at large separations the state in Eq.~\eqref{color3} behaves like the product of two color singlets. There is only one possible orbital, namely $b\bar q$, but the unperturbed state now is the superposition of two states with $\bar q$ bound to one or to the other $b$ 
\be
\Psi(1,2)=\frac{\psi(1)\phi(2)+\phi(1)\psi(2)}{\sqrt{2\left(1+S^2 \right)}}
\ee
The denominator is needed to normalize $\Psi(1,2)$ and it arises because $\psi(1)$ and $\phi(1)$ are not orthogonal, %being eigenfunctions of different hamiltonians, one centered in ${\bm x}_A$ and the other in ${\bm x}_B$.
%\be
%H_{A,B}=\frac{p^2}{2m_q} -\frac{1}{3}\alpha_S\frac{1}{|{\bm x}-{\bm x}_{A,B}|}+k|{\bm x}-{\bm x}_{A,B}|   \label{unpertbb3}
%\ee
with the overlap $S$ defined as
\be
S=\int d^3\xi~\psi(\xi)\phi(\xi)
\ee
%(recall that $\psi(\xi)=R(|{\bm \xi }-{\bm x }_A|)$ and $\phi(\xi)=R(|{\bm \xi }-{\bm x }_B|)$). 
The perturbation Hamiltonian is
\bea
H_{\rm pert}&=&-\frac{1}{3}~\alpha_S \left(\frac{1}{|{\bm x}_1-{\bm x}_B|}+\frac{1}{|{\bm x}_2-{\bm x}_A|}\right) +\nonumber \\
&&-\frac{2}{3}~\alpha_S~\frac{1}{|{\bm x}_1-{\bm x}_2|}
\eea
and
\be
V_{\rm BO}(r_{AB})=2 (\langle H\rangle_{\rm min} +V_0)-\frac{2}{3}\alpha_S\frac{1}{r_{AB}} + \delta E
\ee
where $\delta E=\big( \Psi(1,2), H_{\rm pert}\Psi(1,2)\big)$ evaluates to
\be
\delta E=\frac{1}{1+S^2}\left[ -\frac{2}{3}\alpha_S( I_1+S I_2)-\frac{2}{3}\alpha_S(I_4 + I_6)\right]\label{bb3}
\ee
$I_{1,4}$ were defined previously whereas~\cite{pauling}
\bea
&& I_2(r_{AB})=\int d^3\xi\, \psi(\xi)\phi(\xi)\frac{1}{|{\bm \xi}-{\bm x}_B|} \label{i2}\\
&& I_6(r_{AB})=\int d^3\xi d^3\eta\, \psi(\xi)\phi(\xi) \psi(\eta)\phi(\eta)\frac{1}{|{\bm \xi}-{\bm \eta}|}
\label{i6}
\eea
For the orbital $b\bar q$ we find
$
A=0.44~{\rm GeV},\, \langle H\rangle_{\rm min} = 0.75~{\rm GeV} 
%\label{orbitbb3}    
$.
%@@@@@@@@@@@@@@@@@@@@@@@@@@@@@@@@@@@@@@@@@
\begin{figure}[htb!]
\begin{minipage}[c]{4.2cm}
   \centering
 %\begin{center}
   \includegraphics[width=4.2truecm]{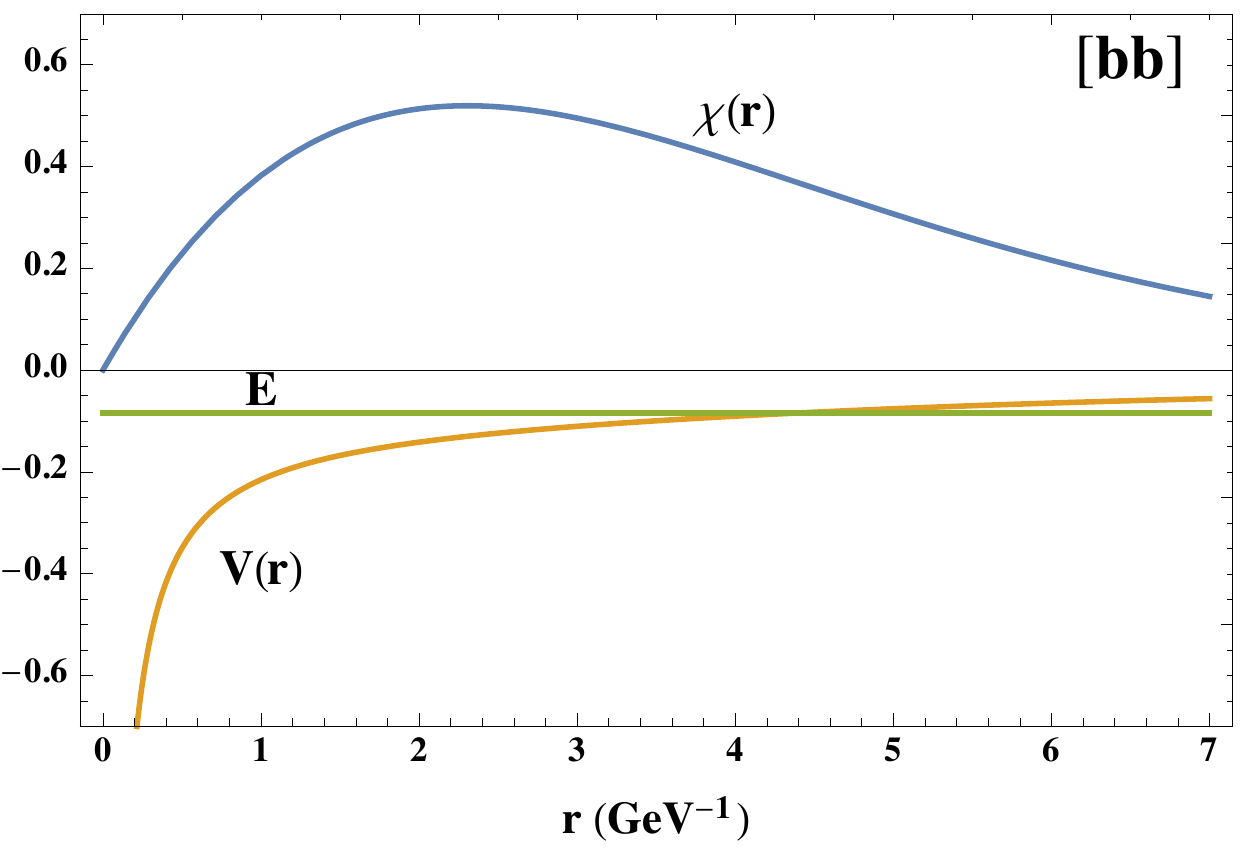}
    \end{minipage}%
% \end{center}
 \begin{minipage}[c]{4.8cm}
\centering
   \includegraphics[width=4.26truecm]{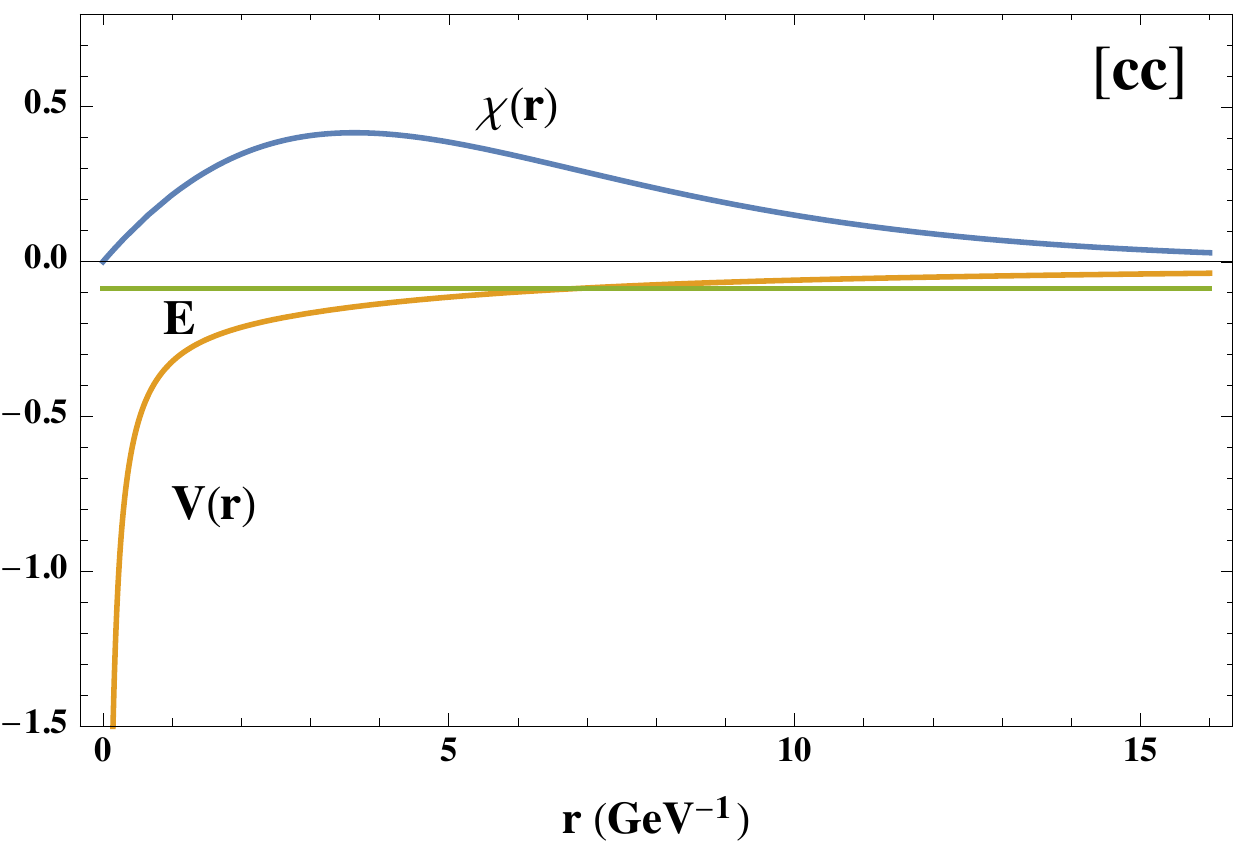}
 \end{minipage}%
\caption{\footnotesize Left Panel: BO potential, eigenfunction and eigenvalue $(bb)_{\bar{\bm 3}}\bar q\bar q$ tetraquark. Right Panel: same for $(cc)_{\bar{\bm 3}}\bar q\bar q$.  \label{qq3}}
\end{figure}
The BO potential, wave function and eigenvalue for the $bb$ pair in color  ${\bar{\bm 3}}$ and the one-gluon exchange couplings are reported in Fig.~\ref{qq3}. There is a bound tetraquark with a tight $bb$ diquark, of the kind expected in the constituent quark model~\cite{Karliner:2017qjm,Eichten:2017ffp,Luo:2017eub}.

The BO potential in the origin is Coulomb-like and it tends to zero, for large $r_{AB}$, due to~\eqref{noconf}. 
The (negative) eigenvalue $E$ of the Schr\"odinger equation is the binding energy associated with the BO potential. The mass of the lowest tetraquark with $(bb)_{S=1},~(\bar q\bar q)_{S=0}$ and of the $B$ mesons are
\bea
&& M(T)=2(M_b + M_q) + E+\frac{1}{2}\kappa_{bb}-\frac{3}{2}\kappa_{qq}\\
&&M(B)=M_b + M_q -\frac{3}{2}\kappa_{b\bar q}
\eea
where $\kappa_{bb}=15$~MeV, $\kappa_{qq}= 98$~MeV and $\kappa_{b\bar q}= 23$~MeV~\cite{Maiani:2004vq} 
are the hyperfine interactions and $E=-84$~MeV is the eigenvalue shown in Fig~\ref{qq3}(a) with  $\alpha_s(m_b)=0.20$. 

\begin{table}[ht!]
\begin{center}
\begin{tabular}{|c|c|c|c|c|}
\hline
$QQ^\prime\bar u\bar d$              & This work & K\&R~\cite{Karliner:2017qjm}  & E\&Q~\cite{Eichten:2017ffp} & Luo {\it et al.}~\cite{Luo:2017eub}   \\
\hline
$cc\bar u\bar d $ & $-10 (+7)$ & $+140$  &$+102$    & $+39$  \\
$cb\bar u\bar d$  & $-73 (-58) $ &  $\sim 0$  &$+83$  & $-108$      \\
$bb\bar u\bar d$  & $-154 (-137)$ & $-170$ & $-121$  & $-75$   \\
\hline
%\hline
%$cc\bar u\bar d $ & $-10$ & $+102$ & $+39$ & $+140$  \\
%$cb\bar u\bar d$  & $-67$ & $+83$ & $-108$&  $\sim 0$   \\
%$bb\bar u\bar d$  & $-154$ & $-121$ & $-75$ & $-170$ \\
%\hline
%FC Copenhagen     & 6 & 2 & 1 & 3 &    \\
\end{tabular}
\end{center}
\caption{\footnotesize $Q$ values in MeV for decays into meson+meson+$\gamma$. The models in~\cite{Karliner:2017qjm,Eichten:2017ffp,Luo:2017eub} are different elaborations of the constituent quark model we use throughout this paper. More details can be found in the original references. We also refer the reader to the lattice QCD literature providing alternate conclusions on these states~\cite{latts}. Results in parentheses are obtained with a string tension $k=1/4\times 0.15$~GeV$^2$ in Eq.~\eqref{orbitpot}. }\label{tab}
\end{table}

The $Q$-value for the decay $T\to 2 B +\gamma$ is then
\be
Q_{bb}=E+\frac{1}{2}\kappa_{bb}-\frac{3}{2}\kappa_{qq}+3~\kappa_{b\bar q}=-154~{\rm MeV}
\label{eqe}
\ee
Results for $Q_{cc,bc}$ are reported in Tab.~\ref{tab} using $\alpha_s((m_b+m_c)/2)=0.23$.
Eq.~\eqref{eqe} underscores the result obtained by Eichten and Quigg~\cite{Eichten:2017ffp} that the $Q$-value goes to a negative constant limit for $M_Q\to \infty$: $Q= -150$~MeV$+{\cal O}(1/M_Q)$. 

{\bf \emph{Double beauty tetraquarks: $\bm b\bm b$ in $\bm 6$.}}
We start from $T=|(bb)_{\bm 6}, (\bar q\bar q)_{{\bar {\bm 6}}}\rangle$, also considered in~\cite{Luo:2017eub}, to find 
\be
T
=\sqrt{\frac{2}{3}}|(\bar q b)_{\bm 1},(\bar q b)_{\bm 1}\rangle _{\bm 1}+\sqrt{\frac{1}{3}}|(\bar q b)_{\bm 8},(\bar q b)_{\bm 8}\rangle _{\bm 1}
\ee
therefore
\be
\lambda_{bb}=\lambda_{\bar q \bar q}=+\frac{1}{3}\alpha_S%\label{repbb}\\
\quad \lambda_{b\bar q}%=\lambda_{q \bar b}
=-\frac{5}{6}\alpha_S\label{attrbqbar}
\ee
The situation is entirely analogous to the $H_2$ molecule, with two identical, repelling light particles. 
For the orbital $b\bar q$, we find $A=0.43~{\rm GeV}$ and $\langle H\rangle_{\rm min}=0.72$~GeV.
The BO potential  with the one-gluon exchange parameters admits a very  shallow  bound state with $E=-32$~MeV, quantum numbers:  $(bb)_{{\bm 6},S=0}$ and $(\bar q\bar q)_{{\bar {\bm 6}},S=0, I=1}$, $J^{PC}=0^{++}$, and charges~$-2,-1,0$.
The $Q$-value for the decay $T\to 2 B$ is then
\be
Q_{bb}=E-\frac{3}{2}\kappa_{bb}-\frac{3}{2}\kappa_{qq}+3~\kappa_{b\bar q}=-70~{\rm MeV}
\ee
 %The picture of diquark-antidiquark states segregated in space by a potential barrier is compatible with the existence of charged partners of the $X^0(3872)$ to be found in $X^\pm\to \rho^\pm\,J/\psi$ final states, with  branching fractions considerably smaller than in the neutral channel.  This requires to push way further on the available experimental bounds.   It also gives an independent thrust to the idea of stable $bb\bar q\bar q$ tetraquarks, still awaiting an experimental {\color{blue}confirmation}. 

We are grateful for hospitality by the T.D. Lee Institute and Shanghai Jiao Tong University where this work was initiated. 
We acknowledge interesting discussions with A. Ali, N. Brambilla,  A. Esposito, R. Lebed and W. Wang.

%@@@@@@@@@@@@@@@@@@@@@@@@@@@@@

\end{document}